%
%
\magnification=1200
\nopagenumbers
\font \tf=cmr10 scaled \magstep2
\def \kopf#1#2#3{ \rightline {KA-TP-#1-#2}
\rightline {#3 #2} \vskip 2.5cm }
\def \titel#1{ {\tf \centerline{#1}} }
\def \tit#1{  {\tf \centerline{#1}} }
\def \ka { {\it
      \centerline {Institut f\"ur Theoretische Physik}
      \centerline {Universit\"at Karlsruhe}
      \centerline {Kaiserstrasse 12, D-76128 Karlsruhe}
      \centerline {Germany} }}
\def \abstract{\centerline {ABSTRACT} \vskip .3cm \rm}
%
%
\tolerance=2000\hbadness=2000
\magnification=1200
\overfullrule=0pt

\def\phi{\varphi}

\def\noin{\noindent}
\def\pn{\par\noindent}
\def\Na{{\bf N}}
\def\N0{{\bf N_{0}}}
\def\psn{\par\smallskip\noindent}
\def\pbn{\par\bigskip\noindent}
\def\ps{\par\smallskip}
\def\pb{\par\bigskip}
\def\={\ =\ }
\def\:={\ :=\ }
\def\+{\ +\ }
\def\-{\ -\ }

\def\Chi{\raise2.5pt\hbox{$\chi$}} 
%
\nopagenumbers {
\kopf {1}{1996}{January}
\titel {The Measure of the Orthogonal Polynomials Related to }
\tit {Fibonacci Chains: The Periodic Case *}
\vfill
\centerline {Wolfdieter L a n g $\ \ ^{**}$ }
\vskip .3cm
\ka
\vskip 4cm
\abstract {\noindent The spectral measure for the two families of
orthogonal polynomial systems related to periodic chains with
N-particle elementary unit  and nearest neighbour harmonic interaction 
is computed using two different methods. The interest is in the orthogonal
polynomials related to Fibonacci chains in the periodic approximation.
The relation of the measure to appropriately defined Green's functions 
is established.\pbn
}

\bigskip
\bigskip
\bigskip
\bigskip
\noindent * \ Herrn Dr. Harald Fried in Dankbarkeit gewidmet
\smallskip
\noindent ** E-mail address: wolfdieter.lang@phys.uni-karlsruhe.de
\vfill
\eject}
%
%
\nopagenumbers
\magnification=\magstep 1
\headline={\hss\tenrm\folio\hss}
\pageno 2
\noindent {\bf 1. Introduction}
\bigskip
Two associated families of orthogonal polynomial systems govern
longitudinal time-stationary vibrations of linear chains with nearest-
neighbour harmonic interaction (springs $\kappa_{n}$, masses $m_{n}$).
\par These polynomials are defined by  three-term
recurrence relations appropriate for orthogonal polynomials [1].
In the case of a mono-atomic chain (mass $m_{0}$) with uniform coupling (strenght 
$\kappa$) they are deformations of Chebyshev's $S_{n}(2(1-x))\equiv 
U_{n}(1-x)$ polynomials of the second kind. $x\equiv \omega^{2}/2\omega_{0}^{2}$,
with $\omega_{0}^{2}=\kappa/m_{0}$, is a normalized frequency-squared.\par
Our interest is in Fibonacci-chains which have uniform coupling ($\kappa_{n}=
\kappa$) and two masses (mass-ratio $r\equiv m_{1}/m_{0}$) 
distributed at site number $n=1,2,...$ in accordance with the binary sequence 
$1,0,1,1,0,1,0,...$ This quasi-periodic sequence is generated by the Fibonacci 
substitution rule $1\to 1,0$ and $0\to 1$.
Such chains have been considered as simple models for special binary 
alloys [2]. The same structure is encountered in the problem of the phonon 
spectrum of a one-dimensional Fibonacci quasicrystal [3]. In this case 
the associated orthogonal polynomial systems are denoted by 
$\{S_{n}^{(r)}(x)\}$ and $\{{\hat S}_{n}^{(r)}(x)\}$ [4]. Up to now their 
spectral measures (or moment functionals) have not been determined. From 
rigorous results on the Fibonacci Hamiltonian in the context of the 
one-dimensional Schr\"odinger equation one expects that this measure is of 
the singular continuous type supported by a Cantor set of zero Lebesgue 
measure [5].
\par As an approximation to the quasiperiodic problem we identify in this
work the spectral measure for {\it periodic} Fibonacci chains with an 
elementary unit consisting of $N$ masses following the pattern of the first $N$ entries
of the above given binary sequence. The measures for general $N$-periodic 
orthogonal polynomial systems  $\{{\cal S}_{n}(x)\}$ and 
$\{\hat {\cal S}_{n}(x)\}$, defined by three-term recursion relations with
$N-$periodic coefficients, can be found using two different methods.
\par The {\it first method} employs the Bloch-Floquet solutions for these periodic
chains and is based on the evaluation of a (judiciously chosen) complex
contour integral. This is a special application of a general method available
for orthogonal polynomials  with asymptotically periodic coefficients in the
recursion relation [6]. For our purpose only the strictly periodic case is
of interest. With this restriction a similar procedure is discussed in ref.[7]. 
The {\it second method} uses the fact that the continued fraction associated
with orthogonal polynomials is the Stieltjes transform [8] of the spectral
measure[1b,9]. For periodic systems this continued fraction can be determined
as fixed-point of a certain M\"obius transformation, and the measure is then
obtained {\it via} the Perron-Stieltjes inversion formula (see e.g.[8,10,11]).
 
\par It turns out that the support of the absolutely continuous part of the
measure (the weight function) is, for fixed $N-$periodic chain parameters 
$\kappa_{n}, m_{n}$, given by $N$, in general disjoint, $x-$intervals
(bands). These are determined from the generalized Chebyshev polynomials of
the first kind ${\cal T}_{N}(x)$, the so-called trace polynomials, by the
condition $\vert {\cal T}_{N}(x)\vert \leq 1$. These bands coincide with the 
support of the spectral density (or density of states) of infinite chains. 
The weight function is, however, not given by the density of states.
In general, a discrete point measure (Dirac $\delta-$function measure) is also present. Each gap between
the $N$ intervals which support the continuous measure may contribute one 
point (which may lie on one of the band boundaries).
\par In the case of Fibonacci chains the limit $N \to \infty$ is supposed to 
correspond to the chains based on the quasi-periodic binary Fibonacci sequence. 
At present this limit is beyond our control.\par
The connection of the absolutely continuous part of the  $\cal S$ and 
$\hat {\cal S}$  measures to the inverse of the imaginary 
part of the diagonal input Green's functions 
for the periodic problem is given in an extra section. The differential 
spectral density (or differential density of states) is recovered, as usual, 
from the imaginary part of the average over the elementary $N-$unit of the 
diagonal (or local) Green's functions.\par

In order to familiarise the reader with our notation we start with the
dynamical equations for the displacements $q_{n}(t)=q_{n}\ exp( i\omega t)$ 
at site number $n$
$$ q_{n+1} \- 2\bigl{(}{{1+k_{n}}\over{2}}\- {{\omega^{2}_{0}}\over{\omega^{2}
_{n}}}x\bigr{)}q_{n}\+k_{n}\ q_{n-1}\= 0 \ ,\ \ n\in {\bf Z} \ \eqno(1.1)$$
with $\omega_{n}^{2}\equiv \kappa_{n}/m_{n}\ $, $k_{n}\equiv \kappa_{n-1}/
\kappa_{n}\ $ and the normalised frequency-squared $x\equiv 
\omega^{2}/(2\omega^{2}_{0})$. The spring between site number $n$ and $n+1$
has strength $\kappa_{n}$. $m_{n}$ is the mass at site number $n$.\pn
This recursion relation can be rewritten in terms of standard associated
polynomials with the help of the transfer matrix method. The displacements
are then given by ({\it cf.}[4a] (2.8),(2.10))
$$ q_{n+1}(x)={\cal S}_{n}(x)\  q_{1}(x)-\hat {\cal S}_{n-1}(x)\   q_{0}(x)
\ \ ,\ \ n \in {\bf N} \eqno(1.2) $$
\noindent with arbitrary inputs $q_{1}(x), q_{0}(x)$.\footnote{*}{\noin We do 
not consider negative site numbers here. See ref.[4b] for the negative $n$ 
case.} 
\noindent This expression is obtained from the transfer matrix $M_{n}$  
$$\eqalignno{
\pmatrix{q_{n+1}\cr q_{n}\cr}\=M_{n}\pmatrix{q_{1}\cr q_{0}\cr}\ ,\  & 
M_{n}:= R_{n}R_{n-1}\cdots R_{1}\ \ ,\ \  &(1.3a)\cr
R_{n}:=\left( \matrix { Y(n)&-k_{n} & \cr 1& 0& } \right)\ \ ,&\  Y_{n}(x) :=
\ 2\bigl{(} {{1+k_{n}}\over{2}}\- {{\omega_{0}^{2}}\over{\omega_{n}^{2}}}
x \bigr{)}\ , \ &(1.3b)\cr} $$
\noin and the result  
$$ M_{n}=\left( \matrix {{\cal S}_{n}&-\hat{\cal S}_{n-1}  \cr 
{\cal S}_{n-1}&-\hat {\cal S}_{n-2} }
\right)\ \ ,\ \ \eqno(1.4) $$
\noindent where the polynomials ${\cal S}_{n}$ and $\hat {\cal S}_{n}$
are defined by the following three-term recurrence relation corresponding
to $M_{n}=R_{n}M_{n-1}$ with input $M_{1}=R_{1}$
$$\eqalignno{{\cal S}_{n} &=Y_{n}(x){\cal S}_{n-1}- k_{n}{\cal S}_{n-2}\ \ \ \ \ \ \ \ \ \ \ \  ,
\ \ \ \  {\cal S}_{-1}=0,
\ \  {\cal S}_{0}=1\ \ \ , & (1.5a) \cr
 \hat{\cal S}_{n} &= Y_{n+1}(x) \hat{\cal S}_{n-1}-k_{n+1} \hat {\cal S}_{n-2}\ \ \ \ ,
 \ \ \ \ \ \ \hat{\cal S}_{-1}
 =0,\ \  \hat {\cal S}_{0}=k_{1}\ \ \ .& (1.5b) \cr} $$
\noin In the case of a {\it mono-atomic} chain ($m_{n}=m$) with equal springs 
($k_{n}=1$) the normalised eigenfrequencies-squared of a finte chain 
with $N$ 
atoms and fixed boundary conditions ($q_{0}\=0\=q_{N+1}$) are given by the zeros 
of  ${\cal S}_{N}(x)\equiv S_{N}(2(1-x))\= U_{N}(1-x)$ , where $U_{N}$ are 
Chebyshev's polynomials of the second kind. These zeros are
$$ x_{k}\equiv x_{k}^{(N)} \= 1\- cos\ {{\pi k}\over{N+1}}\= 2\ sin^{2}  
\ {{\pi k}\over{2(N+1)}} \ \ \ ,\ \  k=1,2,...,N  \eqno(1.6)$$
\noin The displacements for the $k$-th mode are 
$q_{n+1}\= S_{n}(2(1-x_{k}))\ q_{1}$, for $n=1,2,...,N$, with arbitrary , $k$ 
dependent, input $q_{1}$.\par
For an {\it infinite} mono-atomic chain one finds one frequency-squared 
band from the condition $\vert T_{N}(1-x)\vert \leq 1$, where $T_{N}(x)$ 
are Chebyshev's polynomials of the first kind.
The number $N$ of atoms in the unit cell is irrelevant because due to double 
zeros of $T_{N}(1-x)\+(-1)^{k+1}$ for $x=\xi_{k}^{(N-1)}$, $k=1,2,...,N-1$,
the $N-1$ gaps degenerate. The $x-$band is, independently of $N$, 
$B\=[0,2]$. In this case  ${\cal S}_{n}(x)\=$ $\hat{\cal S}_{n}(x)$ $\=
S_{n}(2(1-x))$, which are orthogonal on the intervall $[0,2]$ with weight 
function $$w^{(1)}(x) \= {{2}\over{\pi}}\sqrt{x(2-x)} \ \ \ \ . \eqno(1.7) $$
\noin Therefore there is no discrete part of the measure present in this 
mono-atomic case. \par
The general formula for the differential spectral density per particle (also 
called differential {\it density of states}) for an infinite chain with a unit 
of $N$ atoms repeated periodically is determined from the generalized Chebyshev 
polynomials of the first kind, ${\cal T}_{N}(x) \:= {{1}\over{2}}\bigl{(}
{\cal S}_{N}(x)\-\hat{\cal S}_{N-2}(x)\bigr{)}.$ These are the
trace polynomials ${{1}\over{2}}tr\ M_{N}$. (see eqn.(2.4) and {\it cf.}{[4], 
eqn. (3.11))
$${\cal G}_{N}(x)\= {{1}\over{N\pi}} {{(-1)^{k}\ {\cal T}_{N}^{\prime}(x)} 
\over{\sqrt{1-({\cal T}_{N}(x))^{2}}}}\ \ \ , \eqno(1.8) $$
\noin for $x$ in one of the $N$ bands $B_{k}$, $k=1,2,...,N$, which are
determined by $\vert {\cal T}_{N}(x) \vert\ < \ 1$ and ordered with 
increasing $x$. Otherwise the density vanishes.\pn 
In the mono-atomic case this becomes
$$ G_{N}(x)\= G(x)\= {{1}\over{\pi}} {{1}\over{\sqrt{x(2-x)}}}
\ \ \ , \eqno(1.9)  $$
\noin for $x\in [0,2]$, and it is zero otherwise. The $N-$independence is due 
to the identities involving ordinary Chebyshev polynomials: \pn
$1\- (T_{N}(1-x))^{2}\=x(2-x)(S_{N-1}(2(1-x))^{2}\ $ for
$x\in [0,2]$ and $N\in\Na$, as well as
$$ N S_{N-1}(2(1-x))\= -T_{N}^{\ \ \prime}(1-x)\ \ \ .  \eqno(1.10) $$
\noin $(1.8)$ is, by accident, the weight function for Chebyshev's $T_{N}(1-x)$
polynomials (for $N=0$; for other $N$ the weight is twice this). In the general
case the trace polynomials  $\{{\cal T}_{n}(x)\}$ are no longer orthogonal
(see [4a], p. 5402).\par
Quasiperiodic Fibonacci chains are obtained as special case with 
$k_{n}\equiv 1\ $, $\omega^{2}_{0}/\omega^{2}_{n}\= r^{h(n)}\ $, with
the mass-ratio $r\equiv m_{1}/m_{0}$ and the quasiperiodic binary Fibonacci
sequence
$$ h(n)\=  \lfloor(n+1)/\varphi\rfloor \- \lfloor n/\varphi\rfloor\ \ \ , \ \ \
n\in \N0\ \ . \eqno(1.11)$$
\noin In this case we use for the polynomials ${\cal S}_{n}$ and 
$\hat {\cal S}_{n}$ 
the notation $\{S_{n}^{(r)}(x)\}$ and $\{\hat S_{n}^{(r)}\}$.\pbn
\noindent {\bf 2. Bloch-Floquet solutions and the measure}
\bigskip
\par In this section we describe the computation of the measure with respect to
which the polynomial system $\{{\cal S}_{n}(x)\}$ defined in (1.5a)is 
orthogonal, following the general method valid for the $N$-periodic case 
(more generally for the asymptotically periodic case) described in detail in 
ref. [6], ch.2. The measure for the associated polynomials $\{\hat{{\cal S}}_{n}(x)\}$ 
defined in (1.5b) is obtained in the same way. This method rests on the
properties of the Bloch-Floquet solutions for the $N$-periodic problem. 
Directly connected to these solutions is a map $\rm w(x)$ of the complex plane
which enters the definition of a judiciously chosen complex contour integral.
This integral is evaluated in two different ways {\it i)} and {\it ii)}. The
result will be the orthogonality relation and the measure can be read off.
First, however, the necessary information on the Bloch-Floquet solutions will
be given.\ps
The starting point is the recursion formula for the monic orthogonal
polynomial systems (indicated by a tilde) which describe N-periodic chains 
with $\kappa_{n+N}\=\kappa_{n}$ and $m_{n+N}\=m_{n}$. \footnote{*}{\noin All 
quantities depend on the chosen period $N$, $\vec{\omega^{2}}\equiv 
(\omega^{2}_{0},\omega^{2}_{1},...,\omega^{2}_{N-1})$ and $\vec k\equiv 
(k_{1},k_{2},...,k_{N})$ with $\kappa_{0}=\kappa_{N}$. In the sequel this 
dependence will be suppressed.}
$$\eqalignno{
 \tilde{\cal S}_{n}(x)&=(x-c_{n})\ \tilde {\cal S}_{n-1}(x)
 -d_{n}\ \tilde {\cal S}_{n-2}(x)\ \ ,\ \ & \cr
 \tilde{\cal S}_{-1}&=0\ \ ,\ \ \tilde {\cal S}_{0}=1 \ \,\ \
 & (2.1a)  \cr}$$
\noindent with
$$ c_{n}:={{1+k_{n}}\over{2}}{{\omega^{2}_{n}}\over{\omega^{2}_{0}}} \ \ \ ,\ \
\ \ d_{n}:= {{k_{n}}\over{4}}{{\omega^{2}_{n}}\over{\omega^{2}_{0}}} 
{{\omega^{2}_{n-1}}\over{\omega^{2}_{0}}}\ \ .\ \ \eqno(2.1b) $$
 
\noindent The first-associated monic polynomials 
$\{\tilde {\hat {\cal S}}_{n}(x)\}$ 
satisfy (2.1a) with shifted coefficients   \pn
$\hat c_{n}:=c_{n+1}\ \ ,\ \ \hat d_{n} :=d_{n+1}\ $ and the input 
$\tilde {\hat {\cal S}}_{-1}\ =0$, and $\tilde {\hat{\cal S}}_{0}\ =k_{1}$. 
\par
\noindent Orthogonality with positive definite moment functional is guaranteed 
for all $N\in {\bf N}$ by Favard's theorem, because $d_{n} >0$ and the $c_{n}$ 
are real.\par
The relation between the polynomials (1.5) and the monic ones is given by 
($n\geq 1$)
$$ {\cal S}_{n}(x)= (-2)^{n} \prod_{i=1}^{n} {{\omega^{2}_{0}}\over
{\omega^{2}_{i}}} \ \tilde {\cal S}_{n}(x)\ \ \ \ ,\ \ \ \
\hat{\cal S}_{n}(x)=(-2)^{n}k_{1}\prod_{i=2}^{n+1}{{\omega^{2}_{0}}\over
{\omega^{2}_{i}}} \  \tilde {\hat {\cal S}}_{n}(x)\ \ \  . \eqno(2.2)$$
For the Fibonacci chains with $\kappa_{n}\equiv \kappa$ one uses the $N-$
periodic binary sequence $h^{(N)}(n)$ obtained by repetition of the first $N$
entries of the quasiperiodic sequence $\{h(n)\}$ of (1.11). The original 
polynomials $\{S_{n}^{(r)}(x)\}$ and $\{\hat S_{n}^{(r)}(x)\}$ for 
quasiperiodic Fibonacci chains, are then obtained in the limit $N \to \infty$, 
keeping always $n \leq N$.\par
\noindent In the $N-$periodic case the transfer matrix  $M_{N}$ of the 
elementary unit satisfies: $Det\ M_{N} =\kappa_{0}/\kappa_{N}=+1$. 
From this and $M_{n+N}\ = \ M_{n}
M_{N}\ $ one derives, using  $(1.4)$ 
$$\eqalignno{
 {\cal S}_{n+2N}(x)&=2\ {\cal T}_{N}(x)\ {\cal S}_{n+N}(x)-{\cal S}_{n}(x)
  \ \ , &(2.3a) \cr
  \hat{\cal S}_{n+2N-1}(x)&= 2\ {\cal T}_{N}(x)\ \hat {\cal S}_{n+N-1}(x)-
  \hat {\cal S}_{n-1}(x)\ \ ,\ \ &(2.3b) \cr }$$
\noindent with the trace polynomials
$${\cal T}_{N}(x):={{1}\over {2}}tr\ M_{N} =
{{1}\over {2}}({\cal S}_{N}(x)-\hat{\cal S}_{N-2}(x))\ \ .
\ \  \eqno(2.4) $$
In the {\it Fibonacci case} we use the $N-$periodic binary sequence 
$\{h_{N}(n)\}$ obtained from (1.11) by taking $h_{N}(n)=h(n)$ for 
$n=1,2,...,N$ and defining $h_{N}(n+N)=h_{N}(n)$. In this case the 
polynomials ${\cal S}_{N}$, $\hat{\cal S}_{N-2}$, hence ${\cal T}_{N}$, 
coincide with the corresponding quasiperiodic ones. \pn
(2.3) implies for the general solution  for the  displacements $(1.2)$ 
$$q_{n+2N+1}(x)=2\ {\cal T}_{N}(x)\ q_{n+N+1}(x)-q_{n+1}(x)\ \
. \ \
\eqno(2.5) $$
\noindent The two independent Bloch-Floquet solutions are, for arbitrary
input $q_{0}(x)$ and $q_{1}(x)$,
$$q_{n,\pm}(x):=q_{n+N}(x)-\lambda_{N,\pm}(x)\ q_{n}(x)\ \ ,\ \ \eqno(2.6a) $$
\noindent with
$$\lambda_{N,\pm}(x):= {\cal T}_{N}(x) \pm \sqrt {{({\cal T}_{N}(x))}^{2}
-1}\ \ ,\ \ \ \ \eqno(2.6b) $$
\noindent which are the two eigenvalues of $M_{N}$. With this definition
the periodicity condition (2.5) yields $q_{n+N,\pm}(x)=\lambda_{N,\mp}(x)\ 
q_{n,\pm}(x) \ $ .
\noindent $\lambda_{N,\mp}(x) = exp \ i\beta_{N,\mp}(x)$ for
$x$ values in any of the $N$ intervals (bands) $B_{k}$ ,
$k=1,2,...,N$, defined by \footnote {*}{\noin From (1.5) one finds 
${\cal S}_{n}(0)=1+\kappa_{0}\sum_{i=1}^{n}1/\kappa_{i}\ $, 
${\hat {\cal S}}_{n}(0)=k_{1}+ \kappa_{0} \sum_{i=2}^{n+1} 1/\kappa_{i}\ $. 
Therefore, in the $N-$ periodic case ${\cal T}_{N}(0)=1\ $ holds for all 
chains ($\kappa_{0}=\kappa_{N}$). For the proof of the reality of the zeros
of ${\cal T}_{N}(x)\mp 1$ see ref.[6], lemma 2.2.}
$$ \vert {\cal T}_{N}(x) \vert \leq 1 \ \ ,\ \ \eqno(2.7) $$
\noindent which is necessary for harmonic vibrations.
Therefore,
$$ q_{n,\pm}(x)=(\lambda_{N,\mp}(x))^{{{n}\over {N}} }
\phi_{n}(x)\ \ ,\ \  n \in {\bf {N}_{0}} \ \ ,\ \ \eqno(2.8) $$
\noindent with periodic $\phi_{n+N}(x) =\phi_{n}(x)$ , proving
that (2.6a) defines the Bloch-Floquet solutions. The integrated 
spectral density (or integrated density of states) is a continuous function
given by appropriate branches of the Bloch-Floquet phase 
$\beta_{N}(x) \= cos^{-1} {\cal T}_{N}(x)$ for band values $x$ 
and constant pieces in the $N-1$ gaps between the bands. The differential
spectral density (or differential density of states) (1.7) follows from
this after differentiation. \par
\noindent In order to compute the measure with respect to which the
polynomials  $\{{\cal S}_{n}(x)\}$ are orthogonal one considers (see ref. [6]) 
the following map for $x \in {\bf C}$
$${\rm w}(x)=\bigl ({\lambda_{N}(x)}\bigr )^{{{1}\over {N}} } \ \ ,\ \
\eqno(2.9) $$
\noindent with the sign of the square-root in (2.6b) 
chosen such that for real $x\in B_{k}$,$ \ k=1,...,N$,
$\lambda_{N}(x)$ runs along the
unit circle from $+1$ to $-1$ for odd numbered bands and from $-1$ to $+1$
for even numbered ones. This requires the sign $(-1)^{k+1}$ for $x\in
B_{k}$. $\lambda_{N}(x)$ is real for $x$ in the gaps $G_{k}$,
$k=1,...,N-1$, and we choose the sign of the square-root as $(-1)^{k} $ such 
that  $\vert \lambda_{N}(x)\vert>1$ holds. $x\in(x_{max},\infty)$, with the
maximal band value $x_{max}$, is taken as gap $G_{N}$ with the sign choice 
$(-1)^{N}$. Such a sign choice poduces for $\omega(x)$ of (2.9) a path like 
the one shown for the Fibonacci chain example $N=5,r=2$ in {\it fig. 2}, 
if $x$ runs along the real axis from $0$ to $x_{max}$. 
Finally, in the $z=1/{\rm w}(x)$ plane a closed contour is obtained if 
$x$ runs also backwards from $x_{max}$ to $x=0$. This contour is shown for the 
example in {\it fig.3}. The starting point is at $z=+1$.
In {\it figs.2} (resp. {\it 3}) the $B^{'}_{k}\ $ (resp. $B_{k}^{''}$) and 
$G^{'}_{k}\ $ (resp. $G^{''}_{k}$) labels indicate the images of the bands 
$B_{k}$ and gaps $G_{k}$ under the map ${\rm w}(x)$ of (2.9) 
(resp. $z=1/{\rm w}(x)$). That the map ${\rm w}$ is of importance  
is clear from the Bloch-Floquet solution (2.8).\ps 
For the computation of the measure for the polynomials
$\{S_{n}(x)\}$ it was found in ref.[6] that one should consider the $z$-plane 
contour integral for $m,n=0,1,2,...$ 
$$ I\equiv I_{m,n}:= -{{1}\over {2\pi i}}\int_{\Gamma}
{{{\cal S}_{m}(g(z)) \ q_{n+1,\pm}(g(z))}\over {{\cal S}_{N-1}(g(z))}}
\ g'(z)\  dz \ \ ,\ \ \eqno(2.11) $$
\noindent with $g$ the inverse map to $z(x)=1/{\rm w}(x), {\it viz\ } x=g(z)$,
and the Bloch-Floquet solution $q_{n+1,\pm}(x)$ defined by (2.6a)
and (1.2) with the choice $q_{1}=1$ and $q_{0}=0$. The sign is chosen 
as described above in the definition of ${\rm w}(x)$.
In addition, for $x_{max}<x\leq+\infty$ the sign choice is $(-)^{N}$.
The contour $\Gamma$ (with negative orientation) is the union
$\Gamma=\Gamma_{B^{''}} \cup \Gamma_{G^{''}}$ with
$$ \Gamma_{B^{''}}:=\bigl{\{}z\vert z=exp\ i\theta \ \ ,\ \theta \neq
\pm {{\pi}\over {N}}k\ ,\ k=1,2,...,N-1 \bigr{\}} \eqno(2.12)$$
\noindent and $\Gamma_{G^{''}}$ are the $2(N-1)$ closed curves around
the images $G_{k}^{''}$ of the gaps $G_{k}$. 
For the Fibonacci chain example with $N=5, r=2$ see {\it figs. 1} and {\it 3}.
\noindent For later use note that because $q_{n+1,\pm}(x)$ satisfies
the same recursion relation as ${\cal S}_{n}(x)$ one finds, after
comparison of the $n=-1$ and $n=0$ inputs on both sides, (remember $q_{0}=0$)
$$ q_{n+1,\pm}(x_{i})/q_{1,\pm}(x_{i}) = {\cal S}_{n}(x_{i})\ \ \ , 
\eqno(2.13) $$
\noindent with the zeros $x_{i} $, $i=1,2,...,N-1$, of ${\cal S}_{N-1}$.
\noindent The contour integral is computed in two ways {\it i)} adding the
$\Gamma_{B^{''}}$ and $\Gamma_{G^{''}}$ contribution, both transformed to
the $x-$plane, and {\it ii)} evaluating the $z-$plane contour integral with
the help of the residue theorem. \pn 
In the {\it first step} of {\it i)} one computes the $\Gamma _{B''}$ 
contribution using, for $x\in B_{k}$ with sign choice $(-1)^{k+1}$, 
$$ q_{n+1,\pm}(x) = q_{n+1,\mp}(x) + (1/\lambda_{N,\pm}(x)
- \lambda_{N,\pm}(x))\  {\cal S}_{n}(x)\ \ ,\ \  \eqno(2.14) $$
\noindent following from (2.6a), and with the choice $q_{1}=1 ,$ $q_{0}=0 $
in (1.2) one has $q_{n+1}={\cal S}_{n}$. After some rewriting one               
finds (see {\it appendix A1} for details)
$$ I_{B}=-{{1}\over {2\pi i}} \int_{\Gamma _{B^{''}_{-}}}{{(1/\lambda_{N}(x)
-\lambda_{N}(x))\ {\cal S}_{m}(x)\ {\cal S}_{n}(x)}
\over {{\cal S}_{N-1}(x)}}\ g'(z)\  dz \ \ \ , \ \ \ \eqno(2.15) $$
\noindent with $x=g(z)$, and $B''_{-}$ denotes the lower half of the 
punctured unit circle in the $z-$plane.  This can be restated in the x-plane 
(see {\it fig.1} for the Fibonacci case $N=5, r=2$): 
$$ I_{B}={{1}\over {\pi}} \int_{B}{{{\cal S}_{m}(x)\ {\cal S}_{n}(x)
(-1)^{k+1}\sqrt{1-({\cal T}_{N}(x))^{2}}}\over {{\cal S}_{N-1}(x)}}
\  dx \ \ \ .\ \ \eqno(2.16) $$
\noindent The {\it second step} of the computation {\it i)} is that along
$\Gamma_{G''}$. This integral is again evaluated in $x$-space, such that
the $N-1$ (positively oriented) closed contours around the gaps $G_{k}$ are
picked up ({\it cf. fig.1} for the $N=5$, $r=2$ Fibonacci case). 
${\cal S}_{N-1}(x)$
has a simple zero in each of the $N-1$ gaps. This follows from orthogonality
and a lemma (Ya.L. Geronimus, M. Kac and P. van Moerbeke, Lemma
2.2, page 46 in ref.[6]) about the interlacing of the zeros of ${\cal T}_{N}
\mp1$, which are real, and ${\cal S}_{N-1}$ (and $\hat {\cal S}_{N-1}$). 
Note, that ${\cal S}_{N-1}$ zeros may occur at one of the band boundaries, 
coinciding then with zeros of ${\cal T}_{N}-1$ or ${\cal T}_{N}=1$. 
The residue theorem now yields with the help of (2.13)
$$ I_{G}\ =\ - \int dx \bigr {(} \sum_{k=1}^{N-1}\delta (x-x_{k})
   {{q_{1,(-)^{k}}(x)} \over {{\cal S}_{N-1}^{\prime}(x)}} \bigr {)}\
   {\cal S}_{m}(x)\  {\cal S}_{n}(x)\ \ .\ \eqno(2.17) $$
\noindent with the zeros $x_{k}\equiv \xi_{k}^{(N-1)}$ of ${\cal S}_{N-1}$.
\smallskip
\noindent Computation {\it ii)} of (2.11) is done in the $z-$plane using the
residue theorem. The only possible pole inside $\Gamma$, coming from $g'(z)$, 
occurs for $x\to \infty$, {\it i.e} $z=0$. The zeros of ${\cal S}_{N-1}$ are 
outside of $\Gamma$. Therefore only the large $x$ behaviour of the integrand
is of interest. \par
\noindent For large $x$ the ${\cal S}-$polynomials behave like (see (2.2))
 
$${\cal S}_{m}(x) \sim (-2)^{m}\prod_{i=1}^{m}{{\omega^{2}_{0}}\over{\omega^{2}
_{i}}}\ x^{m}  \ \ .\ \ \eqno(2.18) $$
 
\noindent The asymptotics of $q_{n+1,(-)^{N}}(x)$ cannot be found from
its definition (2.6a) directly. Following ref. [6], it is found  from the 
fomula, based on (2.6a) and (2.5)
 
$$ q_{n+1,+}(x)\ q_{n+1,-}(x)=(q_{n+1+N}(x))^{2}-
q_{n+2N+1}(x)\  q_{n+1}(x)\ \ \ .\ \  \eqno(2.19) $$
 
\noindent The r.h.s. can be rewritten, using the $m-$th
{\it associated} polynomials ${\cal S}_{n}^{(m)}$ defined by
$${\cal S}_{n}^{(m)}(x) = Y_{n+m}(x)\ {\cal S}_{n-1}^{(m)}(x)
-k_{n+m}\ {\cal S}_{n-2}^{(m)}(x)\ \ \ ,\ \ \eqno(2.20) $$
 
\noindent with inputs ${\cal S}_{-1}^{(m)}=0$ and ${\cal S}_{0}^{(m)}=
\prod_{i=1}^{m}k_{i}\ $ for $m\in {\bf N}$ and ${\cal S}^{(0)}_{0}={\cal S}_{0}
=1.$
Because ${\cal S}^{(m)}_{n-m}$ satisfies the recursion relation of $q_{n+1}$, with the two 
independent solutions $q_{n+1}$ and $q_{n+1+N}$, with Wronskian $q_{n+1}\ 
q_{n+N}\- q_{n+1+N}\ q_{n}$ satisfying
 
$$ W(q_{n+1},q_{n+1+N})/\prod_{i=1}^{n}k_{i} \= W(q_{1},q_{N+1})\=
q_{1}q_{N}-q_{0}q_{N+1} \ \ ,\ \
\eqno(2.21) $$
\noindent one finds (for general $q_{1}$ and $q_{0}$)
$$ {\cal S}_{n-m}^{(m)}(x)={{1}\over {W(q_{1},q_{N+1})}}\bigl {\{}
q_{m+N}(x)\  q_{n+1}(x) -q_{m}(x)\ q_{n+1+N}(x) \bigr {\}} \ \ .
\ \ \eqno(2.22) $$
 
\noindent Letting $n \to n+N$ and $m \to n+1$ , one obtains 
the r.h.s. of (2.19):
$$ q_{n+1,+}(x)\ q_{n+1,-}(x)=W(q_{1},q_{N+1})\ 
{\cal S}_{N-1}^{(n+1)}(x)\ \ .\ \  \eqno(2.23) $$
\noindent The asymptotic form of $q_{n+1,(-)^{N}}$ can now be inferred
from the one of $q_{n+1,-(-)^{N}}$ which follows without difficulty from
(2.6a) \footnote{*}{\noin The leading coefficient of 
${\cal S}^{(n+1)}_{N-1}(x)$ is, for $n\in \N0$, $(-2)^{N-1}\prod_{j=1}^{n+1}
k_{j}\ \prod_{i=n+2}^{n+N}\omega^{2}_{0}/\omega^{2}_{i}\ $.}. 
For $q_{1}=1$, $q_{0}=0$ one finds that the leading term for large $x$ is
$${{q_{n+1,(-)^{N}}(x)}\over {{\cal S}_{N-1}(x)}} \sim \prod_{j=1}^{n+1}k_{j}/
((-2)^{n+1} \ \prod_{i=1}^{n+1}{{\omega^{2}_{0}}\over{\omega^{2}_{i}}}\ x^{n+1}) \ \ .
\ \ \eqno(2.24) $$
\noindent The residue theorem can now be applied to the contour integral
(2.11) in the $z-$plane. $g'(z) \sim -C/z^{2}$ for small $z$  due
to $g(z)=x$ and $z=1/w(x) \sim C/x$ for large $x$. The capacity
$C$ drops out in the calculation of the residue for $z=0$.
The result is
$$I_{m,n}=+{{1}\over {2}}{{\omega^{2}_{n+1}}\over{\omega^{2}_{0}}}
\prod_{j=1}^{n+1}k_{j}\  \delta_{n,m} \ \ .\ \ \eqno(2.25) $$
\noindent Combining both ways of calculation {\it i)} and {\it ii)} one ends
up with the normalized measure $d\sigma $ for the ortho{\it normal} polynomials 
with ${\it s}_{0}=1$
$$ {\it s}_{n}(x) :=(-1)^{n}\ \sqrt{{{\omega^{2}_{1}}\over{\omega^{2}_{n+1}}}
{{k_{1}}\over{\prod_{j=1}^{n+1}k_{j}}}} \ {\cal S}_{n}(x)\ \ ,\ \ \eqno(2.26) $$
\noindent where the factor $(-1)^{n}$ has been inserted to guarantee
positive leading coefficient.
$$   \int {\it s}_{m}(x)\ {\it s}_{n}(x)\ d\sigma(x) = \delta_{m,n}
\ \ ,\ \ \ m,n \in \N0\ \  \eqno(2.27) $$
\noindent where
$$\eqalignno{ &d\sigma(x) = {\it w}(x)\  dx -\sum_{k=1}^{N-1}
\delta(x-\xi^{(N-1)}_{k})\ {{2}\over{k_{1}}}{{\omega^{2}_{0}}
\over{\omega^{2}_{1}}}{{q_{1,(-)^{k}}(x)}\over 
{{\cal S}_{N-1}^{\ \prime}(x)}}\ dx\ \ , \ \ & (2.28a) \cr
 &{\it w}(x) ={{1}\over {\pi}}{{2}\over{k_{1}}}{{\omega^{2}_{0}}
 \over{\omega^{2}_{1}}}{{(-1)^{k+1}\sqrt{1-({\cal T}_{N}(x))^{2}}}
\over {{\cal S}_{N-1}(x)}} \ , \  x\in B_{k}\  ,
 k=1,2,...,N,\ \ \ \ & (2.28b) \cr
 &q_{1,(-)^{k}}(x) ={\cal S}_{N}(x)-\lambda_{N,(-)^{k}}(x)\ \
,\ \ k=1,2,...,N-1,\  & (2.28c) \cr
&\lambda_{N,(-)^{k}}(x)  ={\cal T}_{N}(x)
+(-1)^{k}\sqrt{({\cal T}_{N}(x))^{2}-1}\ \ , \ \  x \in G_{k},
 &(2.28d) \cr}$$
\noindent The absolutely continuous part of the measure, ${\it w}(x)\ dx$, 
vanishes outside the $N$ bands $B=\cup B_{k}$. It is non-negative because
the sign of ${\cal S}_{N-1}$ in band $B_{k}$ is $(-1)^{k+1}$, due to the fact 
that ${\cal S}_{N-1}(0)\geq +2$ and the interlacing property of its zeros with
the one of ${\cal T}_{N}\mp 1$. The Dirac-measure lives on the $N-1$ zeros 
$\xi^{(N-1)}_{k}$ of ${\cal S}_{N-1}$. The fact that also this measure is 
non-negative will be discussed in the section 4.
\pb
A similar computation can be performed in order to find the measure
for the associated orthogonal polynomial system $\{\hat {\cal S}_{n}\}$ of
(2.2b). One puts in (2.1) $q_{1}=0$ and $q_{0}=-1$. In the integral (2.11)
one uses $\hat{\cal S}$ instead of $\cal S$ and replaces $q_{n+1,\pm}$ by
$ q_{n+2,\pm}:=\hat{\cal S}_{n+N}-\lambda_{N,\pm}\ \hat {\cal S}_{n}$. 
In place of (2.13) one uses here $q_{n+2,\pm}(\hat x_{i})/q_{2,\pm}(\hat x_{i})
\= \hat {\cal S}_{n}(\hat x_{i})/k_{1}\ $ with the zeros $\hat x_{i}$ of 
$\hat {\cal S}_{N-1}$.
The normalized measure for the ortho{\it normal} polynomials with positive
leading coefficient and ${\hat {\it s}}_{0}=1$
$$ \hat{\it s}_{n}(x)=(-1)^{n}{{1}\over{k_{1}}}
\sqrt{{{\omega^{2}_{2}}\over{\omega^{2}_{n+2}}}{{k_{1}k_{2}}\over{\prod_{j=1}
^{n+2}k_{j}}}}\  \hat {\cal S}_{n}(x) \eqno(2.29) $$
\noindent is then
$$\eqalignno{ d\hat {\sigma}(x)&=\hat {\it w}(x)\  dx - 
\sum^{N-1}_{k=1} \delta(x-\hat \xi_{k}^{(N-1)})\ 
2{{\omega^{2}_{0}}\over{\omega^{2}_{2}}} {{k_{1}}\over{k_{2}}}
{{q_{2,(-)^{k}}(x)/k_{1}}\over {\hat {\cal S}_{N-1}^{\prime}(x)}}\  dx
\ \ \ ,\ \ & (2.30a) \cr
\hat {\it w}(x)&={{2}\over {\pi}}{{\omega^{2}_{0}}\over{\omega^{2}_{2}}}
{{k_{1}}\over{k_{2}}} {{(-1)^{k+1}\sqrt{1-({\cal T}_{N}(x))^{2}}}\over
{\hat {S}_{N-1}(x)}}\ \ \ \ ,\ \ x\in B_{k},\ \  k=1,2,...,N,\ \
&(2.30b) \cr
q_{2,(-)^{k}}(x)&=\hat{\cal S}_{N}(x) - k_{1}\ \lambda_{N,(-)^{k}}(x)\ \ ,
\ \ \ x\in G_{k},\ \ k=1,2,...,N-1 ,\ \ &(2.30c) \cr }$$
 
\noindent where $\lambda_{N,(-)^{k}}$ is given in (2.28d), and 
$\hat \xi_{k}^{(N-1)}$ are the zeros of $\hat {\cal S}_{N-1}(x)$.
\bigskip
\bigskip
\noindent {\bf 3. Continued fraction, Stieltjes inversion formula,
and the measure}
\bigskip
The second method to compute the measure for orthogonal polynomials
with periodic recursion formula coefficients relies on the
fact that the continued fraction accompanying this recursion
formula is the Stieltjes transform [8] of the orthogonality measure [1b,9].
In the periodic case the continued fraction can be given explicitly,
and the measure is then determined with the help of the Perron-Stieltjes
inversion formula [10,11].
\par
\noindent The continued fraction which belongs to the recursion formulae
for the (not necessarily $N$-periodic) associated orthogonal 
polynomials $\{{\cal S}_{n}(x)\}$ and $\{\hat {\cal S}_{n}(x)\}$ 
(see (1.5a) and (1.5b)) is
$$-{{k_{1}\omega^{2}_{1}}\over {2\omega^{2}_{0}}}\chi(x)={{\ \ k_{1} 
\ \ \ \ \ \vert} \over {\vert \ Y_{1}(x)\ \ }}- {{\ \ k_{2}\ \ \ \ \ \vert}
\over {\vert \ Y_{2}(x)\ \ }}-...- {{\ \ k_{n}\ \ \ \ \ \vert}\over
{\vert \  Y_{n}(x)\ \ }}-...\ \ \ ,\ \ \eqno(3.1) $$
\noindent where $Y_{n}(x))$ is given by (1.3b).
The factor $-k_{1}\omega^{2}_{1}/2\omega_{0}^{2}$ has been introduced for later 
convenience. The $n$-th
approximation to this continued fraction is for $n\geq 1$ \footnote {*}
{\noin The $n$-th approximation to the continued J-fraction is for $n\geq 1\ $ 
(see (2.1) and (2.2)) $${{\ \ k_{1}\ \ \ \vert}\over {\vert x-c_{1}}}-
{{\ \ \ d_{2}\ \ \ \vert }\over {\vert x-c_{2}}}-...
-{{\ \ \ d_{n} \ \ \vert}
\over {\vert x-c_{n}}} =\chi_{n}(x)\= {{\tilde {\hat{\cal S}}_{n-1}(x)}
\over{\tilde{\cal S}_{n}(x)}} .$$}
$$-{{k_{1}\omega^{2}_{1}}\over {2\omega_{0}^{2}}}\chi_{n}(x)={{\ \ \ \ k_{1} 
\ \ \ \ \ \vert} \over {\vert \ \ \  Y_{1}(x)\ \ \ }}-{{\ \ \ \ k_{2}\ \ \ \ \vert}
\over {\vert \ \ Y_{2}(x)\ \ }}-... -{{\ \ k_{n}\ \ \ \ \ \vert}\over 
{\vert \ Y_{n}(x)\ \ }} = {{\hat {\cal S}_{n-1}(x)}\over {{\cal S}_{n}(x)}}
\ \ ,\ \ \eqno(3.2) $$
\noindent which follows by induction, using the recursion formulae. If the
continued fraction converges 
$ \chi(x) := \lim_{n \to \infty} \chi_{n}(x)\ \ . $
\noindent A fundamental theorem (see e.g. [9], theorem 2.4, or [1b], p.90)
states that $\chi(x)$ is the Stieltjes transform of  the measure
$$ \chi(x) = \int_{-\infty}^{+\infty}{{d\sigma(t)}
\over {x-t}}\ \, \ \ \ x\not\in supp(d\sigma)\ \ .\ \eqno(3.3) $$
\noindent Here $d\sigma$ is the real, positive, and normalized
measure for the orthogonal ${\cal S}-$ polynomials (1.5a) ({\it cf.} [9],
eqs.(2.1) to (2.5))
$$\int {\cal S}_{n}(x)\ {\cal S}_{m}(x) \ d\sigma (x) = {{\omega^{2}_{n+1}}
\over {\omega^{2}_{1}k_{1}}}\prod_{j=1}^{n+1}k_{j}\ \delta_{n,m}\=
{{m_{1}}\over{m_{n+1}}}\delta_{n,m} \ \ , \ \ m,n \in \N0  \eqno(3.4) $$
\noindent The Perron-Stieltjes inversion formula (see e.g.[9,10]) for a real
measure of bounded variation is
 
$$\sigma(t_{2})-\sigma(t_{1})=-{{1}\over {\pi}}
\lim_{\eta \to +0}\int_{t_{1}}^{t_{2}}Im\ \chi(t+i\eta)\ dt\ \ \ ,
\ \  \eqno(3.5) $$
 
\noindent with $\sigma (t_{k}):={{1}\over {2}} (\sigma (t_{k}+0)+
\sigma(t_{k}-0))$ for $k=1,2$. $\bar \chi(x)=\chi(\bar x)$ for $x\in \bf {C}$,
and $\chi$ is analytic for non-real $x$.
\par
\noindent In the {\it $N-$periodic case} $\chi(x)$
can be calculated explicitly as follows ({\it cf.} [7]). Consider, for
fixed $x$, $N$ and parameters $k_{n}$, $\omega^{2}_{n}$, the map in the 
complex $z-$plane
 
$$J_{N}(x;z)\equiv z'={{\ \ k_{1}\ \ \ \vert}\over {\vert\  Y_{1}(x)}}-
{{\ \ k_{2}\ \ \ \ \vert}\over {\vert\ Y_{2}(x)}}- ...- {{\ \ \ k_{N}\ \ 
\ \ \ \ \ \ \vert} \over {\vert\ \ Y_{N}(x)-z}}\ \ \ , \ \ \eqno(3.6) $$
 
\noindent with $Y_{n}(x)$ given in (1.3b). By
induction, with the help of the recursion formulae, one finds
 
$$ J_{N}(x;z)={{\hat {\cal  S}_{N-2}(x)\ z-\hat {\cal S}_{N-1}(x)}
\over {{\cal S}_{N-1}(x)\ z -{\cal S}_{N}(x)}}\ \ \ ,\ \  \eqno(3.7) $$
 
\noindent which is for real $x$ a $SL(2,\bf {R})$ M\" obius transformation
due to 
 
$$1=Det\ M_{N}(x)=\ -{\cal S}_{N}(x)\ \hat{\cal S}_{N-2}(x) +
\ {\cal S}_{N-1}(x)\  \hat {\cal S}_{N-1}(x)\ \ \ .\ \eqno(3.8) $$
 
\noindent Because of $N-$periodicity
$-k_{1}\omega^{2}_{1}\chi(x)/2\omega^{2}_{0}$ is a fixed point of the map (3.6), 
and can be computed from (3.7) as
$$-{{k_{1}\omega_{1}^{2}}\over {2\omega^{2}_{0}}}\chi_{N,\pm}(x) =
\bigl {\{} {{1}\over {2}}
({\cal S}_{N}(x)+\hat {\cal S}_{N-2}(x))\mp \sqrt {({\cal T}_{N}(x))^{2}-1}
\bigr {\}} / {\cal S}_{N-1}(x)  \ \ \ ,\ \ \ \eqno(3.9) $$
 
\noindent where (3.8) was used, and ${\cal T}_{N}(x)$ is given in (2.4).
For $\vert {\cal T}_{N}(x)\vert<1$, which determines $N$ bands $B_{k}$,
$k=1,2,...,N$, $\ \ \chi_{N,\pm}(x)$ becomes complex. For the gaps between
the $N$ bands, $G_{k}$, $k=1,2,...,N-1$, it is real. See $\it {fig.1}$
for the Fibonacci case $N=5, r=2$. Introducing $\lambda_{N,\pm}(x)$ given in
(2.6b), this is rewritten as
 
$$-{{k_{1}\omega_{1}^{2}}\over {2\omega_{0}^{2}}} \chi_{N,\pm}(x)=\bigl {\{}
{\cal S}_{N}(x)-\lambda_{N,\pm}(x) \bigr {\}}/{\cal S}_{N-1}(x)\ \ .\ \
\eqno(3.10) $$
\noindent $\chi_{N,(-)^{N}}(x)$ is for large $x$ proportional to
$1/x$.\footnote {*}{\noin The $x-$asymptotics cannot be found from (3.10). 
One uses \pn
$({\cal S}_{N}\-\lambda_{N,(-)^{N}})/{\cal S}_{N-1}
\= \hat {\cal S}_{N-1}/({\cal S}_{N}- \lambda_{N,(-)^{N+1}})$ which
is identity (2.23) with $q_{0}=0\ $, $q_{1}=1$, (2.6a) and (1.2).}
The sign choice for $x$ values in the bands and gaps will be given 
below.
\par \noindent The measure can now be determined from the inversion formula
(3.5). The absolutely continuous part of the spectral measure,
$d\sigma_{ac}(x)=w(x)\ dx $, is found from
$$ w(x)={{d}\over {dx}}\sigma_{ac}(x)=-{{1}\over {\pi}}
\lim_{\eta \to +0}\ Im\ \chi(x+i\eta) <\infty \ \ .\ \ \
\eqno(3.11) $$
\noin $\chi(x)\equiv\chi_{N}(x)$ has to be defined such that $\bar \chi(x)=
\chi(\bar x)$ holds for complex $x$. This single-valued function is called 
$\chi(x)$. $w(x)$ lives therefore on
the bands $B_{k}$ and coincides with (2.28b), after the sign of the
square-root ({\it i.e.} of the Riemann sheet) has been chosen such that
$w(x) $ is non-negative. See section 4 for the proof that the sign choice in
(2.28b) leads to positive $w$.\par
\noin The discrete part of the spectral measure (sum of Dirac
$\delta$-functions) originates from the simple poles of $\chi(x)$ 
with their residues determining the height of the jumps of the measure.
$$ d\sigma_{Dirac}(x)=-{{2}\over{k_{1}}}{{\omega^{2}_{0}}\over{\omega^{2}_{1}}}
\sum_{k=1}^{N-1}\delta(x-\xi^{(N-1)}_{k})\ 
{{{\cal S}_{N}(x)-\lambda_{N}(x)}\over {{\cal S}_{N-1}^{\ \prime}(x)}}\ dx
\ , \ \ \eqno(3.12) $$
\noindent where the sign of the square-root in $\lambda_{N}(x) $ of
(2.6b) has been chosen as $(-1)^{k}$ for $x\in G_{k}$, like in  the
calculation leading to (2.28c). This choice will be seen in the next section
to produce positive jumps in $\sigma(x)$ at the zeros
$\xi^{(N-1)}_{k}$ of ${\cal S}_{N-1}$ .
For finite $N$ and parameters $\{k_{n}\}, \{\omega^{2}_{n}\}$ there are no other 
singularities of $\chi(x)$ which could contribute to the singular part of the 
measure ({\it cf.} [12], ch. XIII, p.140). \pn 
The measure for the orthogonal $\{{\hat{\cal S}}_{n}(x)\}$
polynomials can be calculated in a similar fashion  by taking into account
also their first associated polynomials $\{S^{(2)}_{n}(x)\}$, defined in (2.20). 
Note that because of the cyclic property of the trace and $N-$periodicity on has: 
$$\hat {\cal T}_{N}(x):=(\hat{\cal S}_{N}(x)-{\cal S}_{N-2}^{(2)}(x))/2
={{1}\over{2}}tr\ \hat M_{N}\= k_{1} {\cal T}_{N}(x)\  \, \ \eqno(3.13)$$
whith $\hat M_{N}:=R_{N+1}R_{N}\cdots R_{2}\ $. More details are found in 
{\it appendix A.2}.\par
We have thus reproduced the results of the previous section.
\pbn
\noindent {\bf 4. General remarks and Fibonacci chain examples}
\bigskip
We first state a simple conclusion concerning the discrete measure (3.12),
which will show its non-negativity. With the mentioned sign choice the
numerator of the k-th term can be rewritten, with (2.6b), (2.4), and (3.8),
like
$${\cal S}_{N}-\lambda_{N}={{1}\over {2}}\bigl {(}
{\cal S}_{N} +\hat{\cal S}_{N-2} \bigr {)} +\ (-1)^{k+1}\sqrt {
\bigl {(}{{1}\over {2}}({\cal S}_{N}+\hat{\cal S}_{N-2})\bigr {)}^{2}-
{\cal S}_{N-1} \hat {\cal S}_{N-1}}\ . \eqno(4.1) $$
\noindent Evaluated at the zero $x_{k}\equiv \xi^{(N-1)}_{k}$ of
${\cal S}_{N-1}$ this becomes
$$ ({\cal S}_{N}-\lambda_{N})\vert_{x_{k}}=\cases{\Theta\bigl{(}
-({\cal S}_{N}+\hat {\cal S}_{N-2})\vert_{x_{k}}\bigr {)}\ ({\cal S}_{N}
+\hat {\cal S}_{N-2})\vert _{x_{k}} & for k even \cr
\Theta\bigl {(}({\cal S}_{N}+\hat {\cal S}_{N-2})\vert_{x_{k}}\bigr{)}\
({\cal S}_{N}+\hat {\cal S}_{N-2})\vert_{x_{k}} & for k odd \cr}
\ \,\  \eqno(4.2) $$
\noindent with the step function $\Theta (x)$.
Due to (3.8) $({\cal S}_{N}+\hat {\cal S}_{N-2})\vert_{x_{k}}$=$\bigl{(}
({\cal S}_{N}(x_{k}))^{2}-1\bigr{)}/{\cal S}_{N}(x_{k})$. The final result
for the discrete measure is
$$ d\sigma ^{(N,r)}_{Dirac}(x)={{2}\over{k_{1}}}
{{\omega_{0}^{2}}\over{\omega_{1}^{2}}} \sum_{k=1}^{N-1}
\delta(x-\xi^{(N-1)}_{k})\ \Theta\bigl{(}(-)^{k+1}({\cal S}_{N}(x)+
\hat {\cal S}_{N-2}(x))\bigr{)}\
{{{\cal S}_{N}(x)+\hat {\cal S}_{N-2}(x)}
\over{-{\cal S}_{N-1}^{\ \prime}}}\ dx \ , \eqno(4.3) $$
\noindent which is always non-negative because the signum of 
${\cal S}_{N-1}^{\ \prime}(\xi_{k}^{(N-1)})$ is $(-1)^{k}$.
\bigskip
\noindent Two remarks are in order.
\smallskip
\noindent {\it i)} There will be no contribution to the discrete measure from
those zeros of ${\cal S}_{N-1}$ which satisfy $ {\cal T}_{N}(\xi_{k}^{(N-1)})
=\pm1 \ \ . $
\noindent In this event the zero of ${\cal S}_{N-1}$ coincides with one of
the band boundaries, and from the definition of ${\cal T}_{N}$ and (3.8) one
finds ${\cal S}_{N}(\xi^{(N-1)}_{k})$=
$-\hat {\cal S}_{N-2}(\xi_{k}^{(N-1)})=\pm1$.
\smallskip
\noindent {\it ii)} A comment on band degeneracy. A gap $G_{k}$
will disappear whenever ${\cal T}_{N}+(-1)^{k+1}$ has a double zero at, say,
$x_{k}$. Because the zero $\xi^{(N-1)}_{k}$ of ${\cal S}_{N-1}$ lies in
the gap or on one of the adjacent band boundaries one has
$x_{k}=\xi^{(N-1)}_{k}$. For the same reason the k-th zero of
$\hat {\cal S}_{N-1}$ is then also $x_{k}$. Due to (3.8) $\hat {\cal S}_{N-2}
(x_{k})=-1/{\cal S}_{N}(x_{k})$, and therefore
${\cal S}_{N}(x_{k})+1/{\cal S}_{N}(x_{k})=2(-1)^{k}\ $
(definition of ${\cal T}_{N}$).
Thus ${\cal S}_{N}(x_{k})=(-1)^{k}=-\hat {\cal S}_{N-2}(x_{k})\ $, and there
will be no contribution to the discrete part (4.3) of the measure from such
disappearing gaps $G_{k}$. \footnote {*}{
\noindent In the $N-$periodic Fibonacci case an example is N=6 
( $(ABA)^{2}$ chains) where one finds double
zeros of ${\cal T}_{6}+1$ at the zeros of ${\cal T}_{3}$ for $k=1,3,5$. }
\bigskip
\noindent Next, we consider some examples of Fibonacci chains.
\smallskip
\noindent In the $N-$periodic Fibonacci case $k_{n}\equiv 1$ and $Y_{n}(x)
=2(1-r^{h_{N}(n)}x)$, with the $N-$periodic binary sequence $\{h_{N}(n)\}$
obtained by continuing the first $N$ entries of $\{h(n)\}$ given by (1.11)
periodically. All polynomials will now depend on $N$ and the mass-ratio
$r\equiv m_{1}/m_{0}$. We shall use non-script symbols for these polynomials
and the $N,r$ labels will be clear from the context.\psn 
{\it a)} First we check the {\it mono-atomic case } $r=1$. The
results have already been given in the introduction. Remark {\it ii)} applies
for all $N-1$ disappearing gaps. \psn
\noindent {\it b)} Put $N=2,\  r\equiv m_{A}/m_{B}\neq1$,
{\it i.e. AB-chains}. We first show that there is no discrete part (4.3)
of the measure. Here remark {\it i)} applies. The zero of $S_{1}(x)$=
$2(1-rx)$ (see [4a]) is $x_{1}=1/r$, and because
$$ T_{2}(x)=2rx^{2}-2(1+r)x+1\ \ \ ,\ \ S_{2}(x)=4rx^{2}-4(1+r)x+3\ \ \ ,\ \
\ \eqno(4.4)$$
\noindent one has  $S_{2}(1/r)=T_{2}(1/r)$=$-1$=$-\hat S_{0}$.
\footnote {**}{
\noindent This is an example where a zero of ${\cal S}_{N-1}$ coincides with
a band boundary without having band degeneracy (no double zero of ${\cal T}_{N}
\mp 1$). It seems to be a counterexample to one part of the statement
found in [6], lemma 2.2, top of page 47 (the  "if" part).} The weight function
is given by 
$$ w(x)={{2r}\over {\pi}}\sqrt {-x(x-{{1}\over {r}})(x-1)(x-1-
{{1}\over {r}})}\  \ /\vert x-{{1}\over {r}} \vert\ \ ,\ \ \eqno(4.5)$$
\noindent for $x$ in any of the two bands
$$\eqalignno {
r\geq 1\ \ :\ \ \ &B_{1}=[0,{{1}\over {r}}]\ \ ,\ \ B_{2}=
[1,1+{{1}\over {r}}]\ \ \ ,\ \ & (4.6a) \cr
r\leq 1\ \ :\ \ \ &B_{1}=[0,1]\ \ \ ,\ \ B_{2}=[{{1}\over
{r}},1+{{1}\over {r}}]\ \ .\ & (4.6b) \cr}$$
\noin The weight function (2.30) for the associated polynomials 
$\{{\hat S_{n}}(x)\}$ is found to be
$$\hat w\={{1}\over{r}} {{\vert x-{{1}\over{r}}\vert}\over{
\vert x-1\vert}}\ w(x) \ \ \ , \eqno(4.7)$$
\noin for $x$ in the bands (4.6) and zero otherwise.\psn
\noindent {\it c)} Put $N=3,\ r\equiv m_{A}/m_{B} \neq 1$,
{\it i.e.} infinite {\it ABA-chains}.
Now the discrete measure is present because  zeros of $S_{2}$
satisfy $2rx_{\mp} =1+r\mp\sqrt {(r-1)^{2}+r}$ and $S_{3}(x_{\mp})$+$\hat
S_{1}(x_{\mp})$= $2(r-1)x_{\mp}\neq 0$. Therefore
$$ d\sigma_{Dirac}={{4r}\over {\sqrt {(r-1)^{2}+r}}}\cases{
 (r-1)\ x_{-}\ \delta(x-x_{-})\ dx & for $r>1 $\cr
 (1-r)\ x_{+}\ \delta(x-x_{+})\ dx & for $0<r<1$ \cr}\ \ .\ \ \eqno(4.8) $$
\noindent The bands are, depending on $sign(1-r)$,
$$\eqalignno {
 r \geq1:\ \ &B_{1}=[0,{{1}\over {2r}}]\ ,\ \ B_{2}=
 [{{1}\over {2r}}b_{-}(r),{{3}\over {2r}}]\  , \ B_{3}=
 [{{2r+1}\over {2r}},{{1}\over {2r}}b_{+}(r)]\ \ , \ \ & (4.9a)  \cr
 r\leq 1:\ \ &B_{1}=[0,{{1}\over {2r}}b_{-}(r)] \ ,\ \ B_{2}=
 [{{1}\over {2r}},{{2r+1}\over {2r}}]\ \ , \ B_{3}=[{{3}\over
 {2r}},{{1}\over {2r}}b_{+}(r)]\  ,\  &(4.9b)
 \cr}$$
\noindent with $b_{\pm}(r) \equiv r+{{3}\over {2}} \pm
\sqrt {r^{2}-r+{{9}\over {4}} }.$
\par
\noindent The weight function (2.28b) becomes for $x \in B_{k}$ ,
$k=1,2,3$,
$$ w(x)={{2}\over {\pi}} r^{2}{{\sqrt {-x(x-{{2r+1}\over {2r}})
(x-{{3}\over {2r}})(x-{{1}\over {2r}})(x-{{1}\over {2r}}b_{+}(r))
(x-{{1}\over{2r}}b_{+}(r))}}\over {(-1)^{k+1}(x-{{1}\over {2r}}d_{-}(r))
(x-{{1}\over {2r}}d_{+}(r))}}\ \ , \ \ \eqno(4.10) $$
\noindent with $d_{\pm}(r)\equiv 1+r\pm \sqrt{r^{2}-r +1}$.
\smallskip
\noindent {\it d)} Put $N=4, r\equiv m_{A}/m_{B}\neq 1$, {\it i.e. ABAA-chains}.
There is no contribution to the discrete measure because one finds (see [4a])
for the zeros of $S_{3}$, {\it viz} $x_{1}=1/r$, $2rx_{\pm}= 1+r \pm
\sqrt{r^{2}+1}$, $S_{4}(x_{k})$+$\hat S_{2}(x_{k})$=$0$ for
$k=1,\pm$. Like in the $N=3$ case one can give the bands and the weight
function explicitly.
\pbn
\noin {\bf 5. Green's functions and the measure}\footnote{*}{This section
is inspired by a paragraph found in ref.[7] for periodic problems.}\pb
The purpose of this section is twofold: {\it i)\ } to define the Green's functions
of the general $N-$periodic problem which satisfy specific boundary conditions
appropriate to the use of the Bloch-Floquet solutions encountered in section 2.
{\it ii)\ } to find the orthogonality measures computed in this work from these
Green's functions. We shall also give the relation of the differential spectral 
density (or differential density of states) (1.8) to theses Green's functions
which turns out to be the standard one. This should clarify the difference
between the absolutely continuous part of the measure and the spectral 
density.\pn
The Green's functions ${\cal G}_{n,m}(x)$ for the $N-$periodic problem
are  defined for $n,m\in \N0$ by
\footnote{**}{ \noindent The dependence on $N, \{k_{n}\}, \{\omega^{2}_{n}\}$
is suppressed. For the monic polynomials (2.1) one uses
$ (x-c_{n+1})\ \tilde {\cal G}_{n,m}(x)\-d_{n+1}\ 
\tilde {\cal G}_{n-1,m}(x)\-\tilde {\cal G}_{n+1,m}(x)\=\delta_{n,m}\ \ . $ 
The relation $(\prod_{i=1}^{n}{{\omega_{0}^{2}}\over{\omega_{i}^{2}}})\ 
{\tilde {\cal G}}_{n,m}\=(-2)^{m-n+1}(\prod_{i=1}^{m+1}
{{\omega_{0}^{2}}\over{\omega_{i}^{2}}})\ {\cal G}_{n,m}\ $ holds.}

$$ Y_{n+1}(x)\ {\cal G}_{n,m}(x) \-{\cal G}_{n+1,m}(x)\-
k_{n+1}\ {\cal G}_{n-1,m}(x)\=\delta_{n,m}\ \ . \ \ \eqno (5.1) $$
\noin The $Y-$coefficient is defined in (1.3b). The inputs are 
${\cal G}_{-1,m}(x)$ and ${\cal G}_{0,m}(x)$. The Green's functions with 
the proper boundary conditions turn out to be the ones constructed from the
Bloch-Floquet solutions (2.6). This solution is of the type
$$ {\cal G}_{n,m}(x)\= a_{m}(x)\ q_{max(n,m)+1,(-1)^{k}}(x)
\ q_{min(n,m) +1,(-1)^{k+1}}(x)\ \ . \eqno(5.2)$$
\noin The sign choice pertains to gap $G_{k}$, for which 
$q_{n+1,(-1)^{k}} \to 0$ for $x\in G_{k}$ and $n\to \infty$ 
(see (2.8)). The coefficient $a_{m}$ is found from (5.1) putting $n=m$ and 
using the recursion formula for the $q_{n,\pm}$. For general input $q_{1}$ and 
$q_{0}$ one finds with the Wronskian (2.21)
$$\eqalignno{
\prod_{i=1}^{m+1}k_{i}\ {\cal G}_{n,m}(x)\= &{{1}\over{W(q_{1},q_{N+1})\ 
(\lambda_{N,(-1)^{k}}(x) \- \lambda_{N,(-1)^{k+1}}(x))}}\ {\bf \cdot} & \cr
& {\bf \cdot}\ q_{max(n,m)+1,(-1)^{k}}(x) \ \ q_{min(n,m)+1,(-1)^{k+1}}(x) 
\ \ . & (5.3)\cr}$$
\noin This is the Bloch-Floquet Green's function vanishing for $x-$values in
gaps for $n\to\infty$ with fixed $m$ and {\it vice versa}. \par
\noindent Our interest is in the diagonal Green's functions 
${\cal G}_{n,n}(x)$ 
which are in fact $q_{0}$ and $q_{1}$ independent. This is due to identity 
(2.23) which shows that the Wronskian drops out. The final result can be written
in terms of the $m-$th associated polynomials $\{{\cal S}^{(m)}_{n}(x)\}$
defined in (2.20) like
$$ (\prod^{n+1}_{i=1}k_{i})\ {\cal G}_{n,n}(x) \= {\cal S}^{(n+1)}_{N-1}(x)/
\Bigl{(}2\sqrt{\bigl{(}{\cal T}_{N}(x)\bigr {)}^{2}-1}
\Bigr{)}\ \ .\eqno(5.4) $$
\noin The sign of the square root depends on the gap and band number. For
the $k-$th gap, $G_{k}$, it is $(-1)^{k}$, for the $k-$th band,
$B_{k}$ it is $(-1)^{k+1}$ in accordance with the remarks found in
{\it section 2}. In particular, one has for the input quantities
$$\eqalignno{
{\cal G}_{-1,-1}(x)\ =&\  {{1}\over{2}}{\cal S}_{N-1}(x)/
\sqrt{\bigl{(} {\cal T}_{N}(x)\bigr {)}^{2}-1}\ \ & (5.5a) \cr
{\cal G}_{0,0}(x)\ =&\  {{1}\over{2k_{1}}}\hat{{\cal S}}_{N-1}(x)/
\sqrt{\bigl{(}{\cal T}_{N}(x)\bigr {)}^{2}-1}\ \ . & (5.5b) \cr}$$
\noin The imaginary part of these input Green's functions are inversely
related to the weight functions computed in this paper. For 
$x\in B_{k}$, $k=1,2,...,N$, one finds
$$Im \ {\cal G}_{-1,-1}(x)\ :=\ \lim_{\eta\to 0_{+}}\ Im\ {\cal G}_{-1,-1}
(x+i\eta)\ =\ {{1}\over{2}} (-1)^{k} {\cal S}_{N-1}(x)/
\sqrt{1-\bigl{(}{\cal T}_{N}(x)\bigr {)}^{2}}\ \ . \eqno(5.6) $$
\noin In the gaps the imaginary part is zero. This shows that the absolutely
continuous part of the $\{{\cal S}_{n}(x)\}$ measure (the weight function),
which lives on the bands, is essentially the negative inverse of the imaginary
part of the ${\cal G}_{-1,-1}$ Green's function.
$$ -Im\ {\cal G}_{-1,-1}(x)\ =\ \omega^{2}_{0}/(\pi\omega_{1}^{2}w(x))\ \ \ , 
\eqno(5.7) $$
\noin with (2.28b) and $x$ in the bands. Similarly
$$ -Im\ {\cal G}_{0,0}(x)\ =\ \omega_{0}^{2}/(\pi\omega^{2}_{2}k_{2} 
\hat w(x))\ \ \ , \eqno(5.8) $$
\noin with (2.30b).
\noin The average over the elementary $N-$unit of the chain for the diagonal
Green's functions ${\tilde {\cal G}}_{n,n}(x) $, belonging to the monic 
polynomials $\{{\tilde{\cal S}}_{n}(x)\}$, can be computed with the help of the
Christoffel-Darboux identities for the Bloch-Floquet solutions (2.6). These
identies follow from the recursion formula ({\it cf.}refs.[1]), and they are
(remember that $k_{1}k_{2}\cdots k_{N}=1$ in the $N-$periodic case)
$$\eqalignno{
&-2\sum_{n=0}^{N-1}\ 
{{\omega^{2}_{0}}\over{\omega^{2}_{n+1}\ \prod_{i=1}^{n+1}k_{i}}}\ 
q_{n+1,+}(x) \ q_{n+1,-}(x)\= & \cr
& q_{N,+}(x)\ q_{N+1,-}^{\prime}(x) - q_{N+1,+}(x) \ q_{N,-}^{\prime}(x) 
 \+ q_{1,+}(x) \ q_{0,-}^{\prime}(x) - q_{0,+}(x)\ q_{1,-}^{\prime}(x) . 
& (5.9) \cr}$$
\noin There is an alternative version of this identity where the derivative
acts on the left factor and an overall minus sign appears. This happens 
because the confluent Christoffel-Darboux identity implies
$$ q_{N,+}(x)\ q_{N,-}(x) \- 
q_{N,+}(x)\ q_{N+1,-}(x)\- q_{1,+}(x) \ q_{0,-}(x)\ +\ q_{0,+}(x)\ q_{1,-}(x)
\equiv 0\ \ \ ,\ \ \eqno(5.10) $$
which is in fact the Wronskian identity analogous to (2.21) for $W(q_{N+1,+},
q_{N+1,-})$.
\noin The connection between ${\tilde{\cal G}}_{n,n}(x)$, for the monic case,
and ${\cal G}_{n,n}(x) $ is
$${\tilde{\cal G}}_{n,n}(x)\ =\ -2\ {{\omega_{0}^{2}}\over{\omega_{n+1}^{2}}}\ 
{\cal G}_{n,n}(x) \ \ .\ \eqno(5.11) $$
\noin Using (5.4), (2.23) for the $(n+1)-$th associated polynomials, 
identity (5.9), and the definitions (2.6) with (1.2), a lenghty calculation 
shows that for general inputs $q_{0}(x)$ and $q_{1}(x)$
$$\sum_{n=0}^{N-1}\ {\tilde{\cal G}}_{n,n}(x) \ =
\ {\cal T}_{N}^{\ \prime}(x)/ \sqrt{\bigl{(} {\cal T}_{N}(x)
\bigr {)}^{2}-1}\ \ \ , \ \ \eqno(5.12) $$
\noin with the sign convention for gaps and bands mentioned earlier. Therefore,
the imaginary part produces the differential spectral density (or differential
density of states) known from the differentiation of the Bloch-Floquet phase:
$$ {{1}\over{\pi}}\lim_{\eta\to 0_{+}}\ Im\ {{1}\over {N}}\sum_{n=0}^{N-1} 
\ {\tilde{\cal G}}_{n,n}(x+i\eta) \ =\ {\cal G}_{N}(x)\ \ \ , \eqno(5.13)$$  
\noin given by (1.8) for $x$ values in the bands and zero otherwise. This
result corroborates the choice of the Bloch-Floquet Green's functions.
\noin As a by-product we find from (5.12),(5.11), and (5.4) the identity
$$ \sum_{n=0}^{N-1}\ {{\omega^{2}_{0}}\over
{\omega_{n+1}^{2}\prod_{i=1}^{n+1}k_{i}}}\ {\cal S}_{N-1}^{(n+1)}(x)\ 
=\ -{\cal T}_{N}^{\ \prime}(x) \ \ ,\ \eqno(5.14)$$
\noin for the $N-$periodic case of the general associated polynomials (2.20) 
which collapses to the well-known identity (1.10) for Chebyshev polynomials for the
mono-atomic case.
\pbn\pbn \pbn
{\bf \noindent Acknowledgements}
\smallskip
\noindent Thanks go to Dr. A. Anzaldo for pointing out ref. [6] at an early
stage of this investigation. He and Dr. B. Klaiber made helpful remarks
for which the author is grateful. The Green's functions have been discussed
with Mr. D. Walther. The referees made valuable suggestions. One of them
proposed to treat the general periodic case .
\pbn\pbn
\noin{\bf Appendix}\psn
{\bf A1. Derivation of eqn.(2.14)\ } ({\it cf.} Ref.[6])\ps
In the $z-$plane $I_{B}$ is given by (2.11) with $\Gamma=\Gamma_{B''} $
defined by (2.12). The sign choice of $q_{n+1,\pm}$ depends on the band
$B_{k}$, $k=1,2,...,N$, where it is $(-1)^{k+1}$. $q_{0}=0$ and $q_{1}=1$
in (2.6a) with (1.2). The relation (2.14) with the appropriate sign choice
for $\lambda_{N}(x)=\lambda_{N,(-1)^{k+1}}(x)$ is used in order to rewrite
$q_{n+1,\pm}$ of (2.11). The piece with $q_{n+1,\mp}$ is, after a change of
variable, seen to be the negative of the original integral $I_{B}$ with the
relevant $q_{n+1,\pm}$ choice for $x\in B_{k}$. The same change of variable
is used to rewrite the second piece coming from (2.14) as twice the integral
over only half of the contour, namely  over 
$\Gamma_{B_{-}^{''}}$ in the lower half of the $z-$plane.\pbn
{\bf A2. Continued fractions for the $\{{\hat{\cal S}}_{n}\}$ measure 
calculation}\ps
The $\{{\hat{\cal S}}_{n}(x)\}$ recursion relation is given by (1.5b). Their
first associated polynomials are $\{{\cal S}_{n}^{(2)}(x)\}$ defined by (2.20).
The normalised measure $d\hat \sigma$ obeys
$$\int \hat{\cal S}_{n}(x)\ \hat {\cal S}_{m}(x) \ d\hat\sigma (x) = 
{{k_{1}}\over{k_{2}}}\prod_{j=1}^{n+2}k_{j}\ {{\omega^{2}_{n+2}}\over 
{\omega^{2}_{2}}} \delta_{n,m}\=
k_{1}^{2}{{m_{2}}\over{m_{n+2}}}\ \delta_{n,m} \ \ , \ \ m,n \in \N0\ .  
\eqno(A.1) $$
\noin The continued fraction $\hat{\Chi}(x)$ which is related to this measure
like in (3.3) has approximants
$$\hat{\Chi}_{n}(x)\={{{\tilde{\cal S}}_{n-1}^{(2)}(x)}\over
{{\tilde{\cal S}}_{n}^{1}(x)}}\=
-{{2\omega^{2}_{0}}\over {k_{2}\omega^{2}_{2}}}\Bigl{(}{{\ \ \ k_{2} 
\ \ \ \ \ \ \vert} \over {\vert \ \ \ Y_{2}(x)\ \ }}- {{\ \ \ k_{3}\ \ \ \ 
\ \ \vert} \over {\vert\ \ \ Y_{3}(x)\ \ }}-...- {{\ \ k_{n+1}\ \ \ \ \ \ \vert}
\over {\vert \ \ Y_{n+1}(x)\ \ }}-...\ \Bigr{)}\ \ \ . \ \ \eqno(A.2) $$
\noin The tilde quantities are monic polynomials. ${\cal S}^{(1)}_{n}\equiv 
{\hat{\cal S}}_{n}\ $The corresponding M\"obius
transformation is
$$\hat J_{N}(x;z)\equiv z'={{\ \ k_{2}\ \ \ \vert}\over {\vert\  Y_{2}(x)}}-
{{\ \ k_{3}\ \ \ \vert}\over {\vert\  Y_{3}(x)}}- ...- {{\ \ \ k_{N+1}\ \ 
\ \ \ \ \ \ \vert} \over {\vert\ \ Y_{N+1}(x)-z}}\=
{{{\cal S}^{(2)}_{N-2}(x)\ z\- {\cal S}^{(2)}_{N-1}(x)}\over
{{\hat {\cal S}}_{N-1}(x)\ z\-{\hat{\cal S}}_{N}(x)}}\ \ . \eqno(A.3) $$
\noin The fixed point solution can be written like
$$-{{k_{2}\omega_{2}^{2}}\over {2\omega_{0}^{2}}} \hat{\Chi}_{N,\pm}(x)=
\bigl {\{}{\hat {\cal S}}_{N}(x)-k_{1}\lambda_{N,\pm}(x) \bigr {\}}/
{\hat {\cal S}}_{N-1}(x)\ \ ,\ \ \eqno(A.4) $$
\noin where $\hat\lambda_{N,\pm}:={\hat{\cal T}}_{N}(x)
\pm\sqrt{({\hat{\cal T}}_{N})^{2}-k_{1}^{2}}\= k_{1}\lambda_{N,\pm}$ was used which 
follows from (3.13) and (2.6b). The measure $d\hat\sigma$ is then computed 
like in (3.5) from $\hat{\Chi}$. With the definition (2.29) of the 
$\{ {\hat{\it s}}_{n}\}$ polynomials one finds $\int d\hat\sigma\ 
{\hat{\it s}}_{n}(x){\hat{\it s}}_{m}(x)\= \delta_{n,m}\ $ with $d\hat\sigma$
given in (2.30).\pbn
\vfill
\eject
{\bf \noindent References}
\bigskip
\bigskip
\noindent [1a]\phantom{xxxx}G. Szeg\" o: {\it " Orthogonal Polynomials "},
Gordon and Breach, New York, 1939
\smallskip
\noindent [1b]\phantom{xxxx}T.S. Chihara: {\it "An Introduction to
Orthogonal Polynomials "}, Gordon and \break
\phantom{yyyxxxx}Breach, New York, 1978
\smallskip
\noindent [2]\phantom{xxxxx}F. Axel, J.P. Allouche, M. Kleman, M. Mend\`es-
France and J. Peyri\`ere, \break
\phantom{[1]xxxxx}J. Physique Coll. $\underline{47}\ $C3, suppl.7 (1986)
181-186
\smallskip
\noindent [3]\phantom{xxxxx}J.M. Luck and D. Petritis, J. Stat. Phys.
$\underline {42}$ (1986) 289-310
\smallskip
\noindent [4a]\phantom{xxxx}W. Lang, J. Phys.A$\ \underline {25}$ (1992)
5395-5413
\smallskip
\noindent [4b]\phantom{xxxx}W. Lang, "Two Families of Orthogonal Polynomial
Systems Related to Fibonacci \break
\phantom{yyyxxxx}Chains ", pp 429-440 in {\it " Applications of Fibonacci 
Numbers ", Vol.5} \break \phantom{yyyxxxx}eds. G.E. Bergum, A.N. Philippou
and A.F. Horadam, Kluwer Academic Pub-\break
\phantom{yyyxxxx}lishers, Dordrecht, 1993
\smallskip
\noindent [5a]\phantom{xxxx}A. S\" ut\H o, J. Stat. Phys. $ \underline
{56}\ $(1989) 525-531
\smallskip
\noindent [5b]\phantom{xxxx}J. Bellissard, B. Iochum, E. Scoppola and D.
Testard, Comm. Math. Phys.\break
\phantom{yyyxxxx}$\underline{125}\ $(1989) 527-543
\smallskip
\noindent [6]\phantom{xxxxx}W. van Assche:{\ \it " Asymptotics for Orthogonal
Polynomials "}, Lecture Notes in \break
\phantom{yyxxxxx}Mathematics, Vol. 1265, Springer, 1987
\smallskip
\noindent [7]\phantom{xxxxx}S.W. Lovesey, G.I. Watson and D.R. Westhead,
Int. J. Mod. Phys. $ \underline {B5}\ $ (1991) \break
\phantom{yyxxxxx}1313-1346
\smallskip
\noindent [8]\phantom{xxxxx}D.V. Widder: {\it " The Laplace Transform "},
Ch. VIII, Princeton University Press,\break
\phantom{yyxxxxx}1946
\smallskip
\noindent [9]\phantom{xxxxx}R. Askey and M. Ismail: {\it Recurrence
relations, continued fractions and orthogonal\break
\phantom{yyyxxxxx}polynomials "}, Memoirs of the
Am. Math. Soc., Nr. 300, $ \underline {49}$ (1984)1-108
\smallskip
\noindent [10]\phantom {xxxx}A. Wintner: {\it " Spektraltheorie der
unendlichen Matrizen "}, Hirzel, Leipzig, 1929
\smallskip
\noindent [11]\phantom{xxxx}N.I. Akhiezer: {\it " The Classical Moment Problem
"}, Oliver \& Boyd, Edinburgh and \break
\phantom{yyyxxxx}London, 1965
\smallskip
\noindent [12]\phantom {xxxx}M. Reed and B. Simon:{\it "Methods of Modern
Mathematical Physics "}, Vol. IV,\break
\phantom{yyyxxxx}Academic Press, New York, 1978
\smallskip
\vfill
\eject
\end